# 3D Fourier transformation light scattering for reconstructing extend angled resolved light scattering of individual particles

Khoi Phuong Dao[1], KyeoReh Lee[2,3], Yuri Hong[4], Seungwoo Shin[1], Sumin Lee[1], Dong Soo Hwang[4], and YongKeun Park[1,2,3,*]

[1]*Tomocube, Inc., Daejeon 34051, Republic of Korea*
[2]*Department of Physics, Korea Advanced Institute of Science and Technology (KAIST), Daejeon 34141, Republic of Korea*
[3]*KAIST Institute for Health Science and Technology, KAIST, Daejeon, 34141, Republic of Korea*
[3]*Division of Environmental Science and Engineering, POSTECH, Pohang 37673, Republic of Korea*
*\*yk.park@kaist.ac.kr*

**Abstract:** We represent three-dimensional Fourier transform light scattering, a method to reconstruct angle-resolved light scattering (ARLS) with extended angle-range from individual spherical objects. To overcome the angle limitation determined by the physical numerical aperture of an optical system, the optical light fields scattered from a sample are measured with various illumination angles, and then synthesized onto the Ewald Sphere corresponding to the normal illumination in Fourier space by rotating the scattered light signals. The method extends the angle range of the ARLS spectra beyond 90°, beyond the limit of forward optical measurements. Extended scattered light fields in 3D and corresponding ARLS spectra of individual microscopic polystyrene beads, and protein droplets are represented.

## 1. Introduction

Angle-resolved light scattering (ARLS) measurements enable the access to the shape and refractive index (RI) information of scatters and have been widely utilized in various fields, ranging from analytical chemistry, biology, environmental science, and astronomy [1-10]. Recently, Fourier Transform Light Scattering (FTLS) was introduced to effective measure ARLS signals from individual microscopic objects [11]. Exploiting quantitative phase imaging [12], FTLS precisely measure optical field images of a sample, from which ARLS signals are obtained by numerical calculating far-field Fraunhofer diffraction [13]. Due to its single sample analysis capability and high signal-to-noise ratio (SNR), FTLS has been utilized for the study of colloidal clusters [14], light trapping material for solar cells [15], biological cells [16-19] and tissues [20, 21]. However, FTLS has limitations in the range of retrieved ARLS spectrum, determined by the numerical aperture (NA) of an optical imaging system. Recently, sFTLS [22] was developed to extend the angle-range of ARLS up to 90°, which is the representation limitation of arcsine. sFTLS extend the angle-range of ARLS exploiting the principle of synthetic aperture imaging, it is only applicable to a thin 2D-like object such as red blood cells; sFTLS generate significant errors for 3D objects such as microspheres.

Here we present three-dimensional (3D) FTLS, which precisely reconstruct extended-angle ARLS signals from 3D symmetric objects. By holographically synthesizing the FTLS signals obtained with various illuminations regarding the precise vectorial position of each signal's coverage on the Ewald Sphere, the covered range of the constructed scattered far-field is not only extended but the SNR is significantly increased, because the contributions from illuminations at various angles have uncorrelated noise. We also experimentally demonstrate the large-angle ARLS measurements of polystyrene beads at both simulation and experimental condition. Moreover, we applied our method on the investigation of protein droplets, determined the protein concentration of each individual droplet via fitting its unique ARLS spectrum with Mie's theoretical calculation [23].

## 2. Theoretical Principles

### 2.1. Fourier Transform Light Scattering (FTLS)

The FTLS reconstructs ARLS signals by numerically propagating an optical field image to the far-field via Fraunhofer diffraction [11]. Optical field images are obtained with quantitative phase imaging techniques or interferometric microscopy [12, 24]. The propagation of a complex optical field along the $z$-direction can be described by Helmholtz's equation considering the paraxial approximation. The solution of Helmholtz's equation is Fraunhofer diffraction:

$$U_z(x,y) = \frac{1}{i\lambda z}\exp\left[-ik\left(\frac{x^2+y^2}{2z}\right)\right]\iint U_0(x_0,y_0)\exp\left[-ik\frac{xx_0+yy_0}{z}\right]dx_0dy_0, \quad (1)$$

where $U_z(x,y)$ is the complex far-field propagation by a distance of $z$ from the initial field $U_0(x_0,y_0)$, $\lambda$ is the wavelength. The propagated field $U_z$ is calculated by 2D Fourier transformation of an input field $U_0$, as $\tilde{U}_z(q_x,q_y) = FT[U_0(x,y)]$, where $q_x = $

$(2\pi n_m/\lambda)\sin\theta\cos\phi$ and $q_y = (2\pi n_m/\lambda)\sin\theta\sin\phi$ are spatial frequencies along *x*- and *y*-direction. $n_m$ is a RI of a propagative medium, and $\theta$ and $\phi$ are polar and azimuthal angles of the scattered vector, respectively.

The angle range of the ARLS spectrum obtained using FTLS is limited by the numerical aperture (NA) of an optical imaging system as NA = $n_m$sin $\theta_{max}$. The NA-limited maximum angle range corresponds to $|k_{NA}| = 2\pi NA/\lambda$, as shown in Fig. 1A.

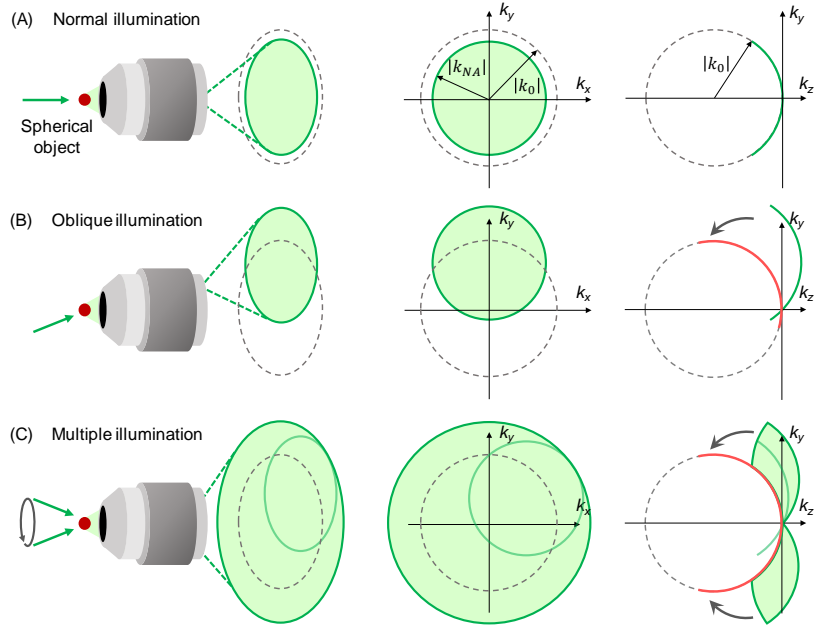

Fig. 1. Schematic of 3D FTLS principle. (A) Illustration of scattered light on the image plane and in Fourier space from a normal illumination, (B) an oblique illumination, and (C) synthesize apertures with various illumination angles. The scattered light in Fourier space is limited by the NA of the objective lens, depicted as filled circle while dot circle depicts the cross-section of the Ewald sphere corresponding to the normal illumination. The rotating and synthesizing mechanism is illustrated in (B) and (C)

### 2.2. Three-dimensional (3D) FTLS

To overcome the angle limitation in FTLS, 3D FTLS synthesizes ARLS signals obtained with various illumination angles, and elastic light scattering signals are mapped onto a spherical surface with the radius $k_0 = 2\pi n_m/\lambda$, called Ewald sphere, in the 3D frequency domain. Each optical field measured with a specific angle of illumination is 2D Fourier transformed, and then mapped onto a corresponding area onto Ewald sphere by rotating the spectra for the origin (Fig. 2B). By synthesizing various optical fields measured with various illumination angles, an ARLS spectrum with the significantly extended angle range is obtained, beyond the ordinarily limited range (Fig. 2C). The 3D FTLS requires a sample to be a point symmetric, such that the rotation of the field information does not change but only expand the angle-range of the ARLS spectra.

From the synthesized ARLS spectra, the scalar field of scattered light, consisting of both the amplitude and phase information, at each point on the Ewald Sphere is paired directly with the corresponding azimuthal and polar angle value. Because 3D FTLS constructs the 3D scattered far-field, instead of projected 2-D far field, the retrieved ARLS spectrum from the mapping process has no distortion at the high-frequency region.

### 2.3. Numerical simulation

To demonstrate the feasibility of 3D FTLS, we performed the numerical simulations with a phantom spherical polystyrene bead with the diameter of 3 μm (*n* = 1.598) immersed in a Norland Optical Adhesive 61 (NOA 61) medium (*n* = 1.565). The bead is exposed to multiple plane illuminations (one normal incident, and 151 azimuthally symmetric incidents with a same tilting polar angle of 44.05°, $\lambda$ = 532 nm), and corresponding scattered optical fields are calculated using the 3D three-dimensional finite-difference-time domain (FDTD), from with 3D FTLS is constructed (Fig. 2). The optical imaging system was assumed to have an NA of 1.2.

From the near field calculated at a monitor plane right after the bead, the FTLS signals were calculated (Fig. 2). From the various illumination angles, the 3D FTLS far-field is constructed (Fig. 2C). The result shows consistency with the analytic solution based on Mie scattering theory. The comparison between Fig. 2a and Fig. 2b demonstrated the extension of the angle range of retrieved ARLS, which results from the larger covered polar-angle range of the rotated Ewald spherical cap. After the rotation and the synthetically stitching process, the angle range of ARLS is significantly expanded. In the ARLS using only recorded field from normal illumination, which is technically FTLS, the maximum observable angle is 67°, and it is fundamentally limited by the NA of the optical system. On the other hand, in 3D FTLS, the angle ranged is extended up to about 110° (Fig. 2C).

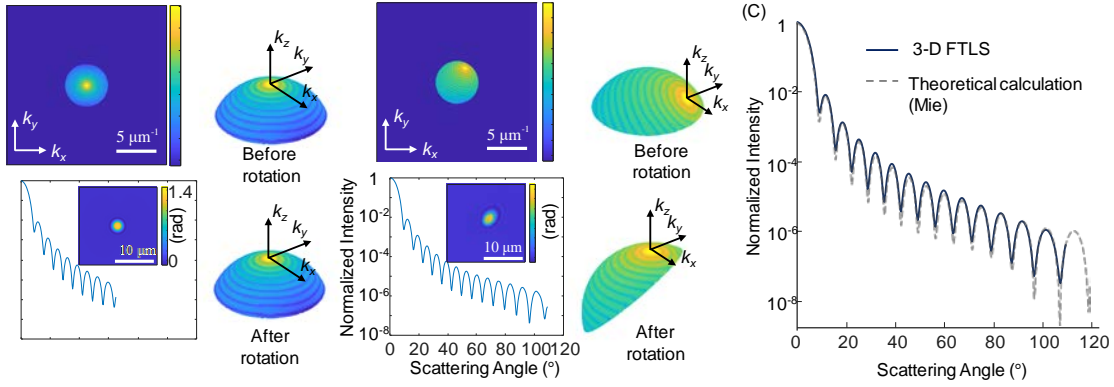

Fig. 2. Numerical simulation of 3D FTLS performance on a polystyrene bead: FTLS signal, mapped FTLS signal on the associated Ewald sphere before and after rotation, and correspondingly constructed ARLS spectrum from a normal illumination (A) and an oblique illumination (B). (C) Synthesized ARLS spectrum.

## 3. Experiments

### 3.1. Experimental setup

An optical diffraction tomography setup (HT-2H, Tomocube Inc., Republic of Korea) was utilized for the experimental validation. The system is based on Mach-Zehnder interferometric microscopy equipped with a digital mirror device (DMD) (Supplement. Fig. S1). Light from a coherent laser source ($\lambda$ = 532 nm, SambaTM, Cobolt Inc., Sweden) was split into a reference and a sample beam using a 1×2 fiber coupler. The DMD (DLPLCR6500EVM, Texas Instruments Inc., USA) systematically controls the angle of the sample beam impinging on a sample by displaying the time-multiplexed hologram patterns [25, 26]. In total, 49 angles (one normal incident, and 48 azimuthally symmetric incidents with a same polar angle of $64.5^0$) controlled using the DMD and a condenser lens (NA = 1.2, water immersion, UPLSAPO 60XW, Olympus Inc.).

The diffracted light from the same is collected using another objective lens (NA = 1.2, water immersion, UPLSAPO 60XW, Olympus Inc.) and projected to the image plane, forming an interference pattern with the reference beam at the camera plane. To maximize the contrast of the pattern, a linear polarizer was placed before the image sensor (FL3-U3–13Y3M-C, Point Grey Research Inc., Canada). The acquisition time for 49 holograms is 0.1 seconds. The processing time for the 3D FTLS was 9 seconds (206 pixels × 206 pixels × 49 illuminations), using a custom MatLab code and a personal computer (Intel Core i7-9700K 3.60 GHz, 32 GB RAM memory).

### 3.2. Inorganic samples: polystyrene beads

To verity the method, we conducted experiments with inorganic beads. Polystyrene beads with the diameter of 5 µm (79633-5ML-F, Sigma-Aldrich Inc.) were in immersion oil ($n_m$ = 1.5, Cat #: 18095, NJ, USA). The results are shown in Fig. 4.

Under the experimental condition, the FTLS angle range is determined to be $\theta_{max} = arcsin(NA/n_m) = 53.1°$ [Eq. (4), Fig. 4(A)]. As the rotation angle in Fourier space of an oblique illumination at $n_m$ = 1.5 is 37.9°, the theoretical angle range of 3D FTLS is extended up to 91.1° (Fig. 3(C)). The averaged ARLS signals over azimuthal angles are presented in Fig. 3(D), and compared with the theoretical calculation based on Mie theory. The conventional FTLS provides the ARLS signals consistent to the theory up to 48° (the red arrow); beyond this angle, the experimental data show errors. Although the sFTLS provides the extended angle range for ARLS signals up to 90°, the measured ARLS signals exhibit significant deviations from the expected theoretical values beyond 22° (the yellow arrow). This is because sFTLS assume a sample as a thin 2D object. The spherical particles cannot be correctly reconstructed by using sFTLS. In contrast, 3D FTLS results exhibit good agreements between the experimental and theoretical data up to 81° (the black arrow). In 3D FTLS, the oscillatory patterns prevail to high angle value. Besides 12 clear picks at an angle lower than 50° that can be easily identified by FTLS, there are 8 picks more at a large angle range (from 50° to 90°) constructed by 3D FTLS.

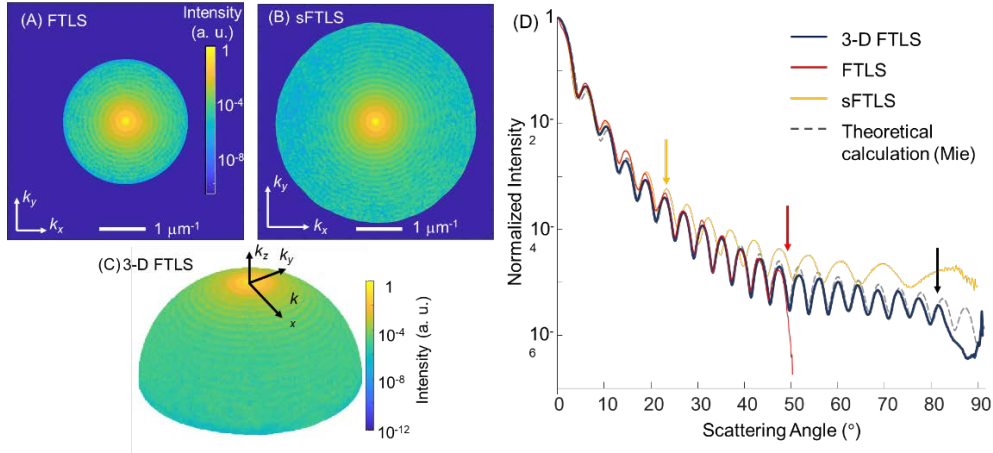

Fig. 3. 3D FTLS performance on Polystyrene bead in experimental condition. Quantitative phase image captured on the image plane, and the associated FTLS signal after the rotation from a normal illumination (A), and an oblique illumination (B). (C) Final scattered field constructed by FTLS, synthetic sFTLS, and 3D FTLS. (D) Corresponding ARLS spectra constructed by FTLS, sFTLS, and 3D FTLS.

### 3.3. Retrieval of physical parameters of microspheres

To precisely retrieve the physical parameters of microspheres, we performed the fitting the 3D FTLS signals of PS microspheres with the Mie theory (Fig. 4). The manufacturer's specifications of the microspheres are the diameter of 2.5 ± 0.1 (STD) µm with the coefficient variation of 2% and the RI of 1.5916.

The 3D FTLS signals of individual microspheres were measured, and then were fitted to Mie scattering patterns with two fitting variables: $\rho_1 = R\, n_m/\lambda$ and $\rho_2 = n/n_m$, where $R$ is the radius of a microsphere, $n$ is the RI of a microsphere, and $n_m$ is the RI of an immersed medium. With the known $n_m$ and $\lambda$, the fitting provides the value of $R$ and $n$ of the microsphere. The fitting was performed using the curve fitting tool in MatLab (R2019b, The MathWorks, Inc.) and Trust-Region algorithm. The result for a representative bead is shown in Fig. 4. The fitting of the 3D FTLS signal to the Mie theory exhibit good matching up to 75°. The retrieved value of $R$ and $n$ of the microsphere are 2.552 µm (confidence bounds: 2.5504 to 2.5531), fitting bead's RI is 1.5790 (confidence bounds: 1.5785 to 1.5795). The retrieved values show good agreements with the manufacturer's specifications: the differences are 0.052 µm (2%) and 0.0126 (0.8%), for $R$ and $n$, respectively.

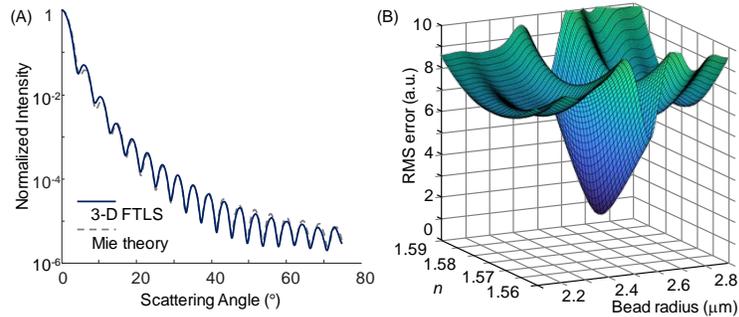

Fig. 4. Precision quantification of the fitting process for (A) 3D FTLS signals and (B) RMS errors in terms of $R$ and $n$.

For statistical analysis, we perform the measurements of 20 PS microspheres. The averaged values of the retrieved the $R$ and $n$ are 2.600 ± 0.060 µm and 1.5761 ± 0.0032, respectively. The difference of the measured $R$ and $n$ with the manufacturer's specifications are 0.1 µm (4%) and 0.0155 (0.9%), respectively. The results show high accuracy and precision.

The measured multiple optical field images of a sample with various illumination angles can also be used for the reconstruction of 3D RI tomogram exploiting the principle of optical diffraction tomography (ODT) [27-29]. From the reconstructed RI tomogram of a spherical particle, one can directly access the information about the radius and RI values. However, the principle of ODT is based on the weak scattering, and thus the retrieved RI values can be under-estimated, especially when the RI difference between a sample and a medium is high. In such cases, there occur multiple internal scatterings which cannot be ignored. We also used ODT and reconstructed the 3D RI tomogram of the same 20 microspheres used in Fig. 4. The retrieved $R$ and $n$ from ODT are 2.79 ± 0.070 µm and 1.5662 ± 0.0011, and the difference with the manufacturer's specification are 0.29 µm (12%) and 0.0254 (1.6%), for *R* and *n*, respectively. The results show that the 3D FTLS provides the values of *R* and *n* of microspheres, more accurately than ODT.

### 3.4. Measurements of protein droplets

To demonstrate applicability of 3D FTLS, ARLS signals of protein droplets were measured. At micrometer-scales, protein or lipid droplets possess spherical-alike morphology and almost uniform density due to the strong surface tension [30]. Characterization of protein droplets are important for the study of phase separation [31, 32], metabolism of cells and related diseases [33, 34], and biotechnology [35, 36]. However, the ARLS signals of individual protein or lipid droplets have not been reported yet.

The ARLS signals of lipid droplets were measured using both the FTLS and 3D FTLS [Fig. 5]. Droplets of protamine protein with various radii ranged from 0.5 to 2.5 µm are immersed in water. The diffracted fields from the sample were recorded as in Section 3.2. The reconstructed RI tomograms of multiple lipid droplets are shown in Fig. 5A. Among them, one representative lipid droplet was selected (the red dotted box in Fig. 5A), from which the mean RI value is retrieved as 1.4177. From the diffracted light field with normal illumination, the FTLS signals is calculated (Fig. 5B). From the diffracted light fields with various illuminations, the 3D FTLS signals are obtained (Fig. 5C). For quantitative analysis, the ARLS signals as a function of a scattering angle were calculated by azimuthally averaging FTLS and 3D FTLS signals (Fig. 5D).

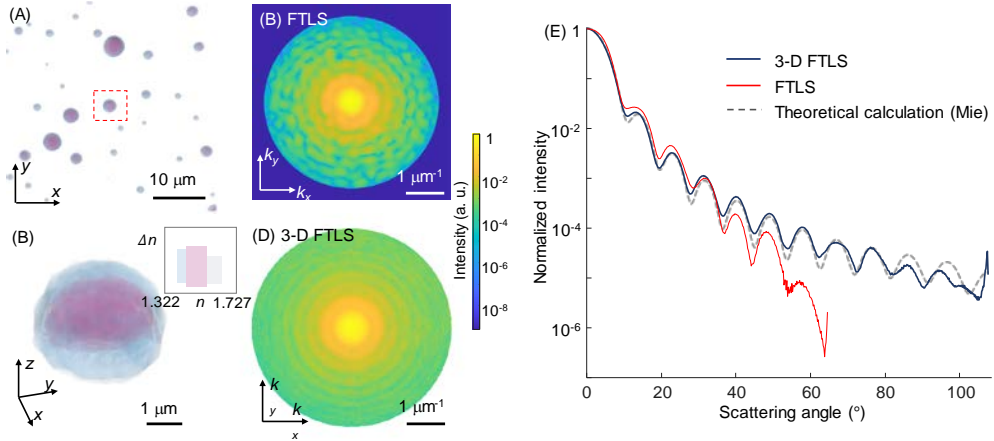

Fig. 5. Experimental results. Synthesized 3D scattered light field and ARLS constructed by 3D FTLS. (A) 3D RI tomogram of protein droplets immersed in water with the selected droplet. (B) Constructed scattered light field by FLTS and 3D FTLS. (C) ARLS spectra of a single protein droplet by FTLS and 3D FTLS.

As shown in Fig. 5, 3D FTLS provide the ARLS signals with the extended angle range up to 108°, whereas FTLS provide the ARLS signals only up to 64°. A most fitted theoretical spectrum based on Mie's theory is also represented for comparison purpose. Mie's scattering signal is calculated with the bead radius of 1.356 µm, the RI of the bead of 1.438, and the RI of the medium of 1.334. the 3D FTLS signals exhibit good agreements to the fitted Mie scattering patterns, with the prominent 11 oscillatory peaks.

## 4. Discussion and conclusions

In sum, we presented 3D FTLS technique, a method to accurately retrieve large-angle ARLS signals of individual particles from multiple optical field measurements with various illumination angles. By rotationally synthesizing each 2D optical scattering signals in Fourier space, ARLS signals of individual microscope objects, including PS beads, PMMA beads, and protein droplets are reported. Compared to the recently developed FTLS or sFTLS techniques, 3D FTLS significantly enhanced the angle range (up to 110°) and SNR.

The present method will find direct applications in the characterizations of microscopic spherical objects such as colloidal particles, and biological liquid droplet. One of the potential applications is the accurate determination of the size and RI for spherical objects. Although 2D QPI or ODT can be used to retrieve the size and RI for spherical objects, the projection of optical phase delay images or the reconstruction of RI tomograms exhibits errors, particularly when the RI contrasts between objects and media are high [37]. Because 3D FTLS provides precise ARLS signals for wide scatting angles with high SNRs, the accurate RI and size information of spherical objects can be retrieved by fitting to Mie theory.

The present method can also be extended to spectroscopic [38-40] or polarization-sensitive ARLS measurements [41], because 3D synthetic process works independently for each wavelength and conserves the polarization sensitivity of scattered light fields. One of the limitations of the present method is the sample should exhibit symmetry, to ensure the rotational synthesis in Fourier space. In the present work, only point-symmetric samples such as spherical particles were measured. However, this condition can be alleviated to rod-like or cylindrical objects [42]. We expect the wealth of information from 3D FTLS would offer unprecedented insight to scientific studies and industrial applications.

## Acknowledgements

This work was supported by KAIST Up program, KAIST Advanced Institute for Science-X, BK21+ program, Tomocube, National Research Foundation of Korea (2017M3C1A3013923, 2015R1A3A2066550, 2018K000396), and KAIST Institute of Technology Value Creation, Industry Liaison Center(G-CORE Project) grant funded by the Ministry of Science and ICT (N11210014). We thank Dr. Seungwoo Shin (KAIST) for helping with FDTD simulation.

## Disclosures

Mr. Dao, Dr. Kyeo Lee, Dr. Sumin Lee, and Prof. Park have financial interests in Tomocube Inc., a company that commercializes optical diffraction tomography and quantitative phase imaging instruments and is one of the sponsors of the work.

## Reference

1. Y. Jo, J. Jung, J. W. Lee, D. Shin, H. Park, K. T. Nam, J.-H. Park, and Y. Park, "Angle-resolved light scattering of individual rod-shaped bacteria based on Fourier transform light scattering," Scientific Reports **4**, 5090 (2014).
2. J. Wang, C. Xu, A. M. Nilsson, D. L. A. Fernandes, and G. A. Niklasson, "A novel phase function describing light scattering of layers containing colloidal nanospheres," Nanoscale **11**, 7404-7413 (2019).
3. P. YongKeun, A. B.-P. Catherine, R. D. Ramachandra, and P. Gabriel, "Light scattering of human red blood cells during metabolic remodeling of the membrane," Journal of Biomedical Optics **16**, 1-7 (2011).
4. M. Fouchier, M. Zerrad, M. Lequime, and C. Amra, "Wide-range wavelength and angle resolved light scattering measurement setup," Optics letters **45**, 2506-2509 (2020).
5. J. W. Pyhtila, and A. Wax, "Rapid, depth-resolved light scattering measurements using Fourier domain, angle-resolved low coherence interferometry," Optics Express **12**, 6178-6183 (2004).
6. N. Yang, W. Angerer, and A. Yodh, "Angle-resolved second-harmonic light scattering from colloidal particles," Physical review letters **87**, 103902 (2001).
7. D. Müller, D. Geiger, J. Stark, and A. Kienle, "Angle-resolved light scattering of single human chromosomes: experiments and simulations," Physics in Medicine & Biology **64**, 045016 (2019).
8. Z. A. Steelman, D. S. Ho, K. K. Chu, and A. Wax, "Light-scattering methods for tissue diagnosis," Optica **6**, 479-489 (2019).
9. A. E. Cannaday, J. Sorrells, and A. J. Berger, "Angularly resolved, finely sampled elastic scattering measurements of single cells: requirements for robust organelle size extractions," Journal of biomedical optics **24**, 086502 (2019).
10. S. Kvaterniuk, V. Pohrebennyk, V. Petruk, O. Kvaterniuk, and A. Kochanek, "Mathematical modeling of light scattering in natural water environments with phytoplankton particles," International Multidisciplinary Scientific GeoConference: SGEM **18**, 545-552 (2018).
11. H. Ding, Z. Wang, F. Nguyen, S. A. Boppart, and G. Popescu, "Fourier transform light scattering of inhomogeneous and dynamic structures," Physical review letters **101**, 238102 (2008).
12. Y. Park, C. Depeursinge, and G. Popescu, "Quantitative phase imaging in biomedicine," Nature Photonics **12**, 578-589 (2018).
13. J. W. Goodman, *Introduction to Fourier optics* (Roberts and Company Publishers, 2005).
14. H. Yu, H. Park, Y. Kim, M. W. Kim, and Y. Park, "Fourier-transform light scattering of individual colloidal clusters," Optics Letters **37**, 2577-2579 (2012).
15. C. Cho, H. Kim, S. Jeong, S.-W. Baek, J.-W. Seo, D. Han, K. Kim, Y. Park, S. Yoo, and J.-Y. Lee, "Random and V-groove texturing for efficient light trapping in organic photovoltaic cells," Solar Energy Materials and Solar Cells **115**, 36-41 (2013).
16. Y. Park, C. A. Best-Popescu, R. R. Dasari, and G. Popescu, "Light scattering of human red blood cells during metabolic remodeling of the membrane," J Biomed Opt **16** (2011).
17. Y. Park, M. Diez-Silva, D. Fu, G. Popescu, W. Choi, I. Barman, S. Suresh, and M. S. Feld, "Static and dynamic light scattering of healthy and malaria-parasite invaded red blood cells," J Biomed Opt **15**, 020506 (2010).
18. Y. Kim, J. M. Higgins, R. R. Dasari, S. Suresh, and Y. K. Park, "Anisotropic light scattering of individual sickle red blood cells," Journal of Biomedical Optics **17**, 040501 (2012).
19. H. Ding, L. J. Millet, M. U. Gillette, and G. Popescu, "Actin-driven cell dynamics probed by Fourier transform light scattering," Biomedical optics express **1**, 260-267 (2010).
20. H. Ding, F. Nguyen, S. A. Boppart, and G. Popescu, "Optical properties of tissues quantified by Fourier-transform light scattering," Optics Letters **34**, 1372-1374 (2009).
21. H. Ding, Z. Wang, F. T. Nguyen, S. A. Boppart, L. J. Millet, M. U. Gillette, J. Liu, M. D. Boppart, and G. Popescu, "Fourier Transform Light Scattering (FTLS) of Cells and Tissues," Journal of Computational and Theoretical Nanoscience **7**, 2501-2511 (2010).
22. K. Lee, H.-D. Kim, K. Kim, Y. Kim, T. R. Hillman, B. Min, and Y. Park, "Synthetic Fourier transform light scattering," Optics Express **21**, 22453-22463 (2013).
23. C. F. Bohren, and D. R. Huffman, "Absorption and Scattering by a Sphere," Absorption and Scattering of Light by Small Particles, 82-129 (1998).
24. K. Kim, and Y. Park, "Fourier transform light scattering angular spectroscopy using digital inline holography," Optics letters **37**, 4161-4163 (2012).

25. S. Shin, K. Kim, J. Yoon, and Y. Park, "Active illumination using a digital micromirror device for quantitative phase imaging," Optics letters **40**, 5407-5410 (2015).
26. K. Lee, K. Kim, G. Kim, S. Shin, and Y. Park, "Time-multiplexed structured illumination using a DMD for optical diffraction tomography," Optics Letters **42**, 999-1002 (2017).
27. K. Kim, J. Yoon, S. Shin, S. Lee, S.-A. Yang, and Y. Park, "Optical diffraction tomography techniques for the study of cell pathophysiology," Journal of Biomedical Photonics & Engineering **2** (2016).
28. E. Wolf, "Three-dimensional structure determination of semi-transparent objects from holographic data," Optics communications **1**, 153-156 (1969).
29. A. Kuś, W. Krauze, P. L. Makowski, and M. Kujawińska, "Holographic tomography: hardware and software solutions for 3D quantitative biomedical imaging," ETRI Journal **41**, 61-72 (2019).
30. S. Alberti, A. Gladfelter, and T. Mittag, "Considerations and Challenges in Studying Liquid-Liquid Phase Separation and Biomolecular Condensates," Cell **176**, 419-434 (2019).
31. Y. Shin, and C. P. Brangwynne, "Liquid phase condensation in cell physiology and disease," Science **357** (2017).
32. S. Boeynaems, S. Alberti, N. L. Fawzi, T. Mittag, M. Polymenidou, F. Rousseau, J. Schymkowitz, J. Shorter, B. Wolozin, and L. Van Den Bosch, "Protein phase separation: a new phase in cell biology," Trends in cell biology **28**, 420-435 (2018).
33. K. Kim, S. Lee, J. Yoon, J. Heo, C. Choi, and Y. Park, "Three-dimensional label-free imaging and quantification of lipid droplets in live hepatocytes," Scientific reports **6**, 1-8 (2016).
34. S. Park, J. W. Ahn, Y. Jo, H.-Y. Kang, H. J. Kim, Y. Cheon, J. W. Kim, Y. Park, S. Lee, and K. Park, "Label-free tomographic imaging of lipid droplets in foam cells for machine-learning-assisted therapeutic evaluation of targeted nanodrugs," ACS nano **14**, 1856-1865 (2020).
35. J. H. Ahn, H. Seo, W. Park, J. Seok, J. A. Lee, W. J. Kim, G. B. Kim, K.-J. Kim, and S. Y. Lee, "Enhanced succinic acid production by Mannheimia employing optimal malate dehydrogenase," Nature communications **11**, 1-12 (2020).
36. U. Schörken, and P. Kempers, "Lipid biotechnology: Industrially relevant production processes," European journal of lipid science and technology **111**, 627-645 (2009).
37. P. Müller, M. Schürmann, S. Girardo, G. Cojoc, and J. Guck, "Accurate evaluation of size and refractive index for spherical objects in quantitative phase imaging," Optics express **26**, 10729-10743 (2018).
38. K. Lee, Y. Kim, J. Jung, H. Ihee, and Y. Park, "Measurements of complex refractive index change of photoactive yellow protein over a wide wavelength range using hyperspectral quantitative phase imaging," Scientific reports **8**, 1-8 (2018).
39. J.-H. Jung, J. Jang, and Y. Park, "Spectro-refractometry of individual microscopic objects using swept-source quantitative phase imaging," Analytical chemistry **85**, 10519-10525 (2013).
40. N. N. Boustany, S. A. Boppart, and V. Backman, "Microscopic Imaging and Spectroscopy with Scattered Light," Annual Review of Biomedical Engineering **12**, 285-314 (2010).
41. J. Jung, J. Kim, M.-K. Seo, and Y. Park, "Measurements of polarization-dependent angle-resolved light scattering from individual microscopic samples using Fourier transform light scattering," Optics Express **26**, 7701-7711 (2018).
42. Q. Zhou, and R. W. Knighton, "Light scattering and form birefringence of parallel cylindrical arrays that represent cellular organelles of the retinal nerve fiber layer," Applied Optics **36**, 2273-2285 (1997).